\documentclass[review]{elsarticle}

\journal{Journal of \LaTeX\ Templates}

\usepackage{numcompress}
\usepackage{amsmath}

\begin{document}

\begin{frontmatter}

\title{Implementation of Composite Photon Blockade Based on Four-wave Mixing System}


\author[mymainaddress,mysecondaryaddress]{Hongyu Lin}

\author[mysecondaryaddress]{Zhi-Hai Yao\corref{mycorrespondingauthor}}
\cortext[mycorrespondingauthor]{Corresponding authors at: Department of Physics, Changchun University of Science and Technology, Changchun 130022, P. R. China.}
\ead{yaozh@cust.edu.cn}

\author[mysecondaryaddress]{Xiao-Qian Wang}

\author[mymainaddress]{Feng Gao}

\address[mymainaddress]{College of Physics and Electronic Information, Baicheng Normal University, Baicheng 137000 P. R. China}
\address[mysecondaryaddress]{Department of Physics, Changchun University of Science and Technology, Changchun 130022, P. R. China}
\begin{abstract}
A high-quality single-photon blockade system can effectively enhance the quality of single-photon sources. Conventional photon blockade(CPB) suffers from low single-photon purity and high requirements for system nonlinearity, while unconventional photon blockade(UPB) has the disadvantage of low brightness. Recent research by [Laser Photon.Rev 14,1900279,2020] demonstrates that UPB can be used to enhance the strength of CPB, thereby improving the purity of single-photon sources. Research by [Opt. Express 30(12),21787,2022] shows that there is an intersection point between CPB and UPB in certain nonlinear systems, where the performance of single photons is better. In this study, we investigated the phenomenon of photon blockade in a non-degenerate four-wave mixing system, where CPB and UPB can occur simultaneously within the same parameter range. We refer to this phenomenon as composite photon blockade. Particularly, when the system achieves composite photon blockade, the value of $g^{(2)}(0)$ is smaller, and there are more single photons. We conducted analytical analysis and numerical calculations to study the conditions for the realization of CPB, UPB, and 2PB in the system, and discussed in detail the influence of system parameters on various blockade effects.
\end{abstract}

\begin{keyword}
CPB\sep 
 UPB\sep 
 Four-wave mixing system.
\end{keyword}

\end{frontmatter}


\section{Introduction}
The development of single-photon source technology is a key driving force behind advancements in quantum communication\cite{1}, quantum metrology\cite{2}, and quantum information technology\cite{3,4}. The primary objective in achieving a single-photon source is to realize non-classical states of light. The study of non-classical photon states is also of profound significance for the development of quantum optics, quantum integration technology, and the future of quantum light fields\cite{5,6}. There are primarily two methods for preparing single-photon sources: weak light pulse attenuation and parametric conversion based on nonlinear materials. However, the weak light pulse attenuation method suffers from high multiphoton probabilities. In contrast, the parametric conversion method relies on achieving photon anti-bunching phenomena. Photon anti-bunching can be realized through two mechanisms: one based on energy level splitting in the system, which primarily relies on strong second-order ($\chi^2$) or third-order ($\chi^3$) nonlinearities. This mechanism is referred to as conventional photon blockade (CPB). The other mechanism is based on destructive quantum interference between different photon state pathways, known as unconventional photon blockade (UPB). In 1997, Imam$\bar{o}$glu et al. first observed the photon blockade effect (PB) in experiments involving optical cavity coupling and trapped atoms\cite{7}. This research breakthrough is considered a milestone in the fields of quantum optics and laser science.\par

The realization of CPB primarily depends on the quantum nonlinear resources within the system. These resources can be attained through high-order nonlinear effects, necessitating a system with a high nonlinear electric polarizability, such as Kerr nonlinear systems\cite{8,9}. Strong coupling between an optical microcavity and atoms within it, known as quantum truncation\cite{10,11,12}, can also achieve these resources. Various experimental schemes have been proposed for realizing quantum nonlinear resources, including optical lattices\cite{13}, superconducting circuits\cite{14}, and ultracold atom trapping\cite{15,16}, among others. CPB has been observed in different optical systems, such as optical microcavities\cite{17,18,19,20,21}, systems embedding quantum dots in resonant cavities of photonic crystals\cite{22}, QED systems\cite{23}, optomechanical systems\cite{24,25,26}, and waveguide systems\cite{27}. Ester et al. achieved single-photon emission in an InGaAs/GaAs quantum dot system. According to their research, when the second-order correlation function ($g^{(2)}(0)$) of the single-photon emission system is less than 0.07, the efficiency of quantum computation can reach 0.9\cite{28}. Therefore, achieving a smaller $g^{(2)}(0)$ value in the single-photon emission system is crucial for excellent single-photon sources. In contrast to CPB, UPB can be realized without relying on the nonlinear intensity of the system. Liew and Savona were the first to achieve PB in weakly nonlinear systems~\cite{29}. The research by Carmichael and Bamba confirmed that PB realized in weakly nonlinear systems primarily relies on destructive quantum interference between different photon state paths\cite{30,31}. This type of PB, based on destructive quantum interference, is also referred to as UPB. Subsequently, Flayac et al. extensively discussed the input-output theory of UPB in a coupled system of two nanocavities\cite{32}. In the same year, Zhou et al. successfully demonstrated UPB in a typical two-mode system and provided detailed discussions on implementation conditions and parameter settings\cite{33}. Currently, UPB has been experimentally verified in various weakly nonlinear optical systems\cite{34,35,36,37,38,39,40,41}.\par

Achieving high-quality photon blockade can effectively drive the development of high-quality non-Poissonian light sources. In certain specific nonlinear optical systems, CPB and UPB not only coexist but also exhibit better single-photon performance at the intersection points\cite{ren,39}. In this work, we discuss CPB, UPB, UCPB, and 2PB in a non-degenerate four-wave mixing system. The results indicate that there is also an intersection point between CPB and UPB in this system, which we refer to as composite photon blockade. Our recent research shows that a four-wave mixing system with Kerr nonlinearity can achieve CPB\cite{wang}, while a four-wave mixing system with two-level atoms can simultaneously achieve CPB and 2PB\cite{yang}. The model discussed in this paper has a simpler nonlinear coupling relationship, making it more advantageous for experimental implementation. Through analytical calculations, we obtained the conditions for achieving CPB and UPB in the system. Numerical results show that the implementation of CPB is consistent with the analytical solution and strengthens with the increase of the nonlinear coefficient. When the conditions for UPB are met, the minimum value of $g^{(2)}(0)$ perfectly matches the analytical solution. Particularly, at the intersection point of the analytical conditions for CPB and UPB, the single photons exhibit better performance, i.e., a smaller $g^{(2)}(0)$ value and a higher average photon number. The discussion on the realization of two-photon blockade (2PB) in the system indicates that the system can achieve 2PB in a single driving mode. Furthermore, we discussed in detail the impact of system parameters on various blockade effects.\par

The manuscript is organized as follows: In Sec.~{\rm II}, we introduced the system model and conducted analytical calculations on the system implementation of CPB and UPB, respectively. In Sec.~{\rm III}, we conducted numerical calculations on the system's implementation of PB. Conclusions are given in Sec.~{\rm IV}.\par

\section{Physical model and analytical solution for implementing PB}
Here, we consider a general three-mode optical system with four-wave mixing interactions. The system's Hamiltonian is given by:
\begin{eqnarray}
\hat{H}&=&\hbar\omega_a\hat{a}^{\dag}\hat{a}\ +\hbar\omega_b\hat{b}^{\dag}\hat{b}+\hbar\omega_c\hat{c}^{\dag}\hat{c} +\hbar g(\hat{a}^2\hat{b}^{\dag}\hat{c}^{\dag}+\hat{a}^{\dag2}\hat{b}\hat{c})\nonumber\\
&&+\hbar F_a(\hat{a}^{\dag}e^{{-i\omega_l}t}+\hat{a}e^{{i\omega_l}t}) +\hbar E(\hat{b}^{\dag}\hat{c}e^{-i(\omega_{l_1}+\omega_{l_2})t}+\hat{c}^{\dag}\hat{b}e^{i(\omega_{l_1}+\omega_{l_2})t}).
\label{01}
\end{eqnarray}
Here, $\hat{a}(\hat{a}^{\dag})$, $\hat{b}(\hat{b}^{\dag})$, and $\hat{c}(\hat{c}^{\dag})$ represent the annihilation (creation) operators of modes $a$, $b$, and $c$, respectively. The parameter $g$ represents the coefficient of nonlinear interactions. $F_a$ represents the driving coefficient of photons in mode $a$, with a driving frequency of $\omega_l$. $E$ represents the non-degenerate parametric amplification coefficient\cite{42,43,44,45} for modes $b$ and $c$, with a frequency of $\omega_{l_1} + \omega_{l_2}$, satisfying the relationship $2\omega_l = \omega_{l_1} + \omega_{l_2}$. (Setting $\hbar=1$) By using the rotating operator $\hat{U}(t) = e^{it(\omega_l\hat{a}^{\dag}\hat{a} + \omega_{l_1}\hat{b}^{\dag}\hat{b} + \omega_{l_2}\hat{c}^{\dag}\hat{c})}$, we can obtain the effective Hamiltonian of the system as:

\begin{eqnarray}
\hat{H}_{eff}&=&\Delta_a\hat{a}^{\dag}\hat{a}\
+\Delta_b\hat{b}^{\dag}\hat{b}+\Delta_c\hat{c}^{\dag}\hat{c}+g(\hat{a}^2\hat{b}^{\dag}\hat{c}^{\dag}+\hat{a}^{\dag2}\hat{b}\hat{c})\nonumber\\
&&+F_a(\hat{a}^{\dag}+\hat{a})+E(\hat{b}^{\dag}\hat{c}+\hat{c}^{\dag}\hat{b}).
\label{02}
\end{eqnarray}
Here, the detuning amounts are given by $\Delta_a = \omega_a - \omega_l$, $\Delta_b = \omega_b - \omega_{l_1}$, and $\Delta_c = \omega_c - \omega_{l_2}$, satisfying the relationship $\Delta_b + \Delta_c = 2\Delta_a$. The Fock-state basis of the system is denoted by $|mnp\rangle$, where $m$ represents the photon number in mode $a$, $n$ represents the photon number in mode $b$, and $p$ represents the photon number in mode $c$.

\subsection{Analytical analysis of CPB in the system}
According to the single excitation resonance condition, when a system resonantly absorbs a photon with frequency $\omega_{l}$, it transitions from the photon state $|0-\rangle$ to the photon state $|1-\rangle$. However, if the system exhibits strong nonlinearity, leading to a detuning between the photon states $|1-\rangle$ and $|2-\rangle$, it can no longer absorb photons of the same frequency to activate the photon state $|2-\rangle$. As a result, the system can only emit an existing photon before the next photon can enter, resulting in the phenomenon of photon anti-bunching and enabling single-photon blockade. Figure. 1 illustrates this concept. In the current system, which has a strong nonlinear coupling strength and a single-mode driving frequency $\omega_{l}=\omega_{a}$, it will not occupy the photon state $|2-\rangle$, thus realizing single-photon blockade. Therefore, the condition for achieving CPB in this system is $\omega_{l}=\omega_{a}$, which corresponds to $\Delta_a=0$.

\begin{figure}[ht]
\centering
\includegraphics[scale=0.60]{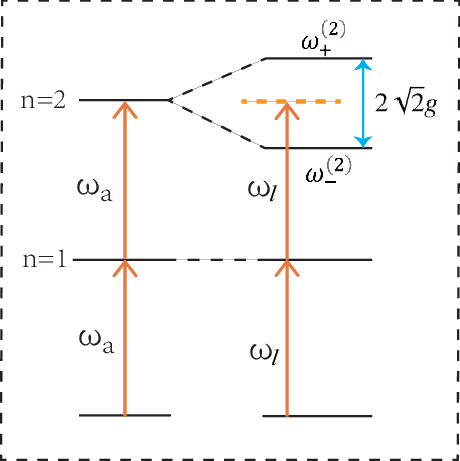}
\caption{Energy level diagram of CPB realized by the system.
} \label{fig1}
\end{figure}

The splitting height of energy level 2 can be obtained by solving the eigenvalues of the system. Considering the photon blockade in mode a, we select $|200\rangle$ and $|011\rangle$ to form a closed subspace.The Hamiltonian can be expanded using $|200\rangle$ and $|011\rangle$, and can be represented in matrix form as:

\begin{eqnarray}
\tilde{H}=
\begin{bmatrix}
2\Delta_a & \sqrt{2}g\\
 \sqrt{2}g & \Delta_b+\Delta_c
\end{bmatrix}, \label{03}
\end{eqnarray}

In fact, the occurrence of photon blockade aims to convert Poisson-distributed photons in the input system into sub-Poisson-distributed photons in the output. When the system achieves photon blockade, it typically requires satisfying the condition of weak driving. Here, by neglecting the driving terms, we can analyze the matrix mentioned above and consider the coupling relationships of photons in each mode. This allows us to obtain the eigenfrequencies of the system, denoted as $\omega_{+}^{(1)}$ and $\omega_{-}^{(1)}$, given by:
\begin{eqnarray}
\omega^{(1)}_{\pm}=2\Delta_a\pm\sqrt{2}g.
\label{04}
\end{eqnarray}
Therefore, when the system realizes photon blockade, the splitting height of energy level 2 is 2$\sqrt{2}g$.

\subsection{Analytical analysis of UPB in the system}
The implementation of UPB differs from CPB in that it relies on destructive quantum interference between different paths. Specifically, the system exhibits at least two paths from the photon state $|1-\rangle$ to the photon state $|2-\rangle$. Destructive interference among these paths prevents the photon state $|2-\rangle$ from being reached, resulting in interference cancellation and allowing only single photon states to exist in the system.
\begin{figure}[ht]
\centering
\includegraphics[scale=0.60]{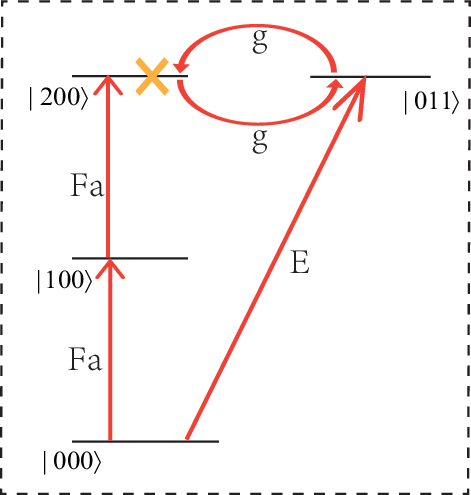}
\caption{The photon state and interference path of UPB are realized in the system.
} \label{fig2}
\end{figure}
The energy level structure and transition paths are illustrated in Figure 2, depicting three paths for the system to reach the two-photon state of mode $a$:
(i) $|000\rangle\stackrel{\underrightarrow{F_a}}{}|100\rangle\stackrel{\underrightarrow{F_a}}{}|200\rangle$.
(ii) $|000\rangle\stackrel{\underrightarrow{E}}{}|011\rangle\stackrel{\underrightarrow{F_a}}{}|200\rangle$.
(iii)$|000\rangle\stackrel{\underrightarrow{F_a}}{}|100\rangle\stackrel{\underrightarrow{F_a}}{}|200\rangle\stackrel{\underrightarrow{g}}{}|011\rangle\stackrel{\underrightarrow{g}}{}|200\rangle$.
When the optimal conditions for photon antibunching are met, photons from different paths destructively interfere at the two-photon state of mode $a$, preventing the occupation of state $|200\rangle$. This constitutes the physical mechanism underlying the implementation of UPB. The system's wavefunction can be expressed as:
\begin{eqnarray}
|\psi\rangle &=& C_{000}|000\rangle+C_{100}|100\rangle+C_{200}|200\rangle+C_{011}|011\rangle.
\label{05}
\end{eqnarray}
This approximation holds when the system contains only a single photon in mode $a$. Taking into account environmental effects, we can treat the system using a non-Hermitian Hamiltonian:
\begin{eqnarray}
\widetilde{H}=\hat{H}_{\text{eff}}-i\frac{\kappa_a}{2}\hat{a}^\dag\hat{a}
-i\frac{\kappa_b}{2}\hat{b}^\dag\hat{b}-i\frac{\kappa_c}{2}\hat{c}^\dag\hat{c}.
\label{06}
\end{eqnarray}
Here, $\kappa_a$, $\kappa_b$, and $\kappa_c$ represent the decay rates of cavity $a$, $b$, and $c$, respectively. For convenience in calculations, we set $\kappa_a=\kappa_b=\kappa_c=\kappa$. The effective Hamiltonian $\hat{H}_{\text{eff}}$ is given by Equation (\ref{02}).
By substituting the wavefunction (from Equation (\ref{05})) and the non-Hermitian Hamiltonian into the Schr\"{o}dinger equation $i\partial_t|\psi\rangle=\widetilde{H}|\psi\rangle$, we obtain a set of coupled equations for the different coefficients:
\begin{eqnarray}
&&i\dot{C}_{000}=F_a C_{100}+E C_{011},\nonumber\\
&&i\dot{C}_{100}=F_a C_{000}+(\Delta_a-\frac{i\kappa}{2})C_{100}+\sqrt{2}F_a C_{200},\nonumber\\
&&i\dot{C}_{200}=\sqrt{2}F_a C_{100}+2(\Delta_a-\frac{i\kappa}{2})C_{200}+\sqrt{2}g C_{011},\nonumber\\
&&i\dot{C}_{011}=E C_{000}+ \sqrt{2}g C_{200}+(\Delta_b-\frac{i\kappa}{2})C_{011}+(\Delta_b-\frac{i\kappa}{2})C_{011}.
\label{07}
\end{eqnarray}
Thus, the steady-state coefficient equations can be expressed as:
\begin{eqnarray}
&&F_a C_{000}+(\Delta_a-\frac{i\kappa}{2})C_{100}+\sqrt{2}F_a C_{200}=0,\nonumber\\
&&\sqrt{2}F_a C_{100}+2(\Delta_a-\frac{i\kappa}{2})C_{200}+\sqrt{2}g C_{011}=0,\nonumber\\
&&E C_{000}+\sqrt{2}g C_{200}+(\Delta_b-\frac{i\kappa}{2})C_{011}+(\Delta_b-\frac{i\kappa}{2})C_{011}=0.
\label{8}
\end{eqnarray}
In general, the probability amplitudes satisfy the relation $|C_{000}|\gg |C_{100}|\gg |C_{200}|$. Therefore, for convenience in calculations, we can set $|C_{200}|=0$and $|C_{000}|=1$. Additionally, the detunings between different photon modes satisfy $\Delta_a=\Delta_b+\Delta_c$. Under the aforementioned conditions, we solve Equation (\ref{8}) and obtain the following solution:
\begin{eqnarray}
E = -\frac{2 F_a^{2}}{g},
\label{9}
\end{eqnarray}
which represents the optimal analytical condition for achieving UPB in mode $a$.

Based on the analytical analysis presented earlier regarding the implementation of CPB and UPB, we have derived the optimal conditions for achieving both CPB and UPB. In the case of a driving frequency $\omega_a=\omega_l$, corresponding to $\Delta_a=0$, the system can achieve CPB. Conversely, UPB can be realized under the condition described by Equation (\ref{9}). Interestingly, we can now explore whether the system can achieve a stronger form of single-photon blockade known as composite photon blockade (PB) while simultaneously satisfying the conditions for CPB and UPB. Below, we will perform numerical calculations to analyze the implementation of PB in the system.

\section{The numerical results for the photon blockade}
In non-classical optics, we can determine the statistical properties of mode $a$ photons by numerically solving the correlation function $g_a^{(n)}(0)$. If $g_a^{(n)}(0)<1$, it indicates a low probability of $n$ photons being present simultaneously, implying the presence of anti-bunching effect in mode $a$ photons. Therefore, when $g_a^{(2)}(0)<1$, it signifies sub-Poissonian statistics of mode $a$ photons, with a small probability of two-photon occupancy, indicating single-photon blocking in mode $a$. To achieve two-photon blocking, both three-photon anti-bunching and two-photon bunching conditions should be simultaneously satisfied, i.e., $g_a^{(2)}(0)\geq1$ and $g_a^{(3)}(0)<1$. The brightness of the system can be represented by the average photon number $N=\langle \hat{a}^{\dag}\hat{a} \rangle=\text{Tr}(\hat{\rho}_s\hat{a}^{\dag}\hat{a})$. To realize a single-photon source, it is necessary to ensure a strong anti-bunching effect in the system, along with a sufficiently high brightness. The expression for the correlation function $g_a^{(n)}(0)$ can be obtained by solving the master equation. The dynamics of the current system's density matrix $\hat{\rho}$ can be described by the following equation:
\begin{eqnarray}
\frac{\partial\hat{\rho}}{\partial t}&=&-i[\hat{H},\rho]+\frac{\kappa_a}{2}(\bar{n}_{ta}+1)(2\hat{a}\hat{\rho}\hat{a}^\dag+\frac{1}{2}\hat{a}^\dag\hat{a}\hat{\rho}+\frac{1}{2}\hat{\rho}\hat{a}^\dag\hat{a})\nonumber\\
&&+\frac{\kappa_b}{2}(\bar{n}_{tb}+1)(2{b}{\rho}{b}^\dag+\frac{1}{2}{b}^\dag{b}\hat{\rho}+\frac{1}{2}{\rho}{b}^\dag{b})\nonumber\\
&&+\frac{\kappa_c}{2}(\bar{n}_{tc}+1)(2\hat{c}\hat{\rho}\hat{c}^\dag+\frac{1}{2}\hat{c}^\dag\hat{c}\hat{\rho}+\frac{1}{2}{\rho}\hat{c}^\dag{c})\nonumber\\
&&+\frac{\kappa_a}{2}\bar{n}_{ta}(2\hat{a}\hat{\rho}\hat{a}^\dag+\frac{1}{2}\hat{a}^\dag\hat{a}\hat{\rho}+\frac{1}{2}\hat{\rho}\hat{a}^\dag\hat{a})\nonumber\\
&&+\frac{\kappa_b}{2}\bar{n}_{tb}(2\hat{b}\hat{\rho}\hat{b}^\dag+\frac{1}{2}\hat{b}^\dag\hat{b}\hat{\rho}+\frac{1}{2}\hat{\rho}\hat{b}^\dag\hat{b})\nonumber\\
&&+\frac{\kappa_c}{2}\bar{n}_{tc}(2\hat{c}\hat{\rho}\hat{c}^\dag+\frac{1}{2}\hat{c}^\dag\hat{c}\hat{\rho}+\frac{1}{2}\hat{\rho}\hat{c}^\dag\hat{c}).
\label{08}
\end{eqnarray}
Where $\kappa_a$, $\kappa_b$, and $\kappa_c$ represent the decay rates of modes $a$, $b$, and $c$, respectively. $\bar{n}{ta}$, $\bar{n}{tb}$, and $\bar{n}{tc}$ represent the average number of thermal photons in modes $a$, $b$, and $c$, respectively. Their expressions are given by $\bar{n}{ti}={\exp{[\hbar\omega_i/(\kappa_BT_i)-1]}}^{-1}$, where $\kappa_B$ is the Boltzmann constant and $T_i$ is the temperature of the reservoir in thermal equilibrium. In the subsequent numerical discussions, for convenience, we assume $\bar{n}{ta} = \bar{n}{tb}=\bar{n}{tc} = n{th}$. Here, we only need to consider the steady-state values of $g_{a}^{(n)}(0)$. Therefore, we can simply set $\partial\hat{\rho}/\partial t=0$ to solve the master equation for the steady-state density operator $\hat{\rho}_s$. The statistical properties of photons will be described using the zero-delay time correlation function:
\begin{eqnarray}
g_a^{(n)}(0)=\frac{\langle
\hat{a}^{\dag n}\hat{a}^n\rangle}
{\langle \hat{a}^{\dag}\hat{a}\rangle^n}.
\label{005}
\end{eqnarray}
By numerically solving the master equation, we can obtain the values of $g^{(n)}(0)$. Smaller values of $g^{(n)}(0)$ indicate a strong photon bunching effect occurring in the system, which is an essential condition for realizing an ideal single-photon source.

\subsection{Numerical Results of the system in implementing CPB}

Next, we will further investigate the scenarios of achieving conventional photon blockade (CPB), unconventional photon blockade (UPB), composite photon blockade (UCPB), and two-photon blockade (2PB) in the system through numerical calculations. We will also compare the numerical results with analytical results. The numerical calculations primarily involve plotting the correlation function $g_a^{(n)}(0)$ and the average photon number $N_a$ of mode $a$ as functions of the system parameters. All numerical calculations for modes $a$, $b$, and $c$ are performed in truncated Hilbert spaces of dimension 5. For convenience, we will redefine all parameters in units of the dissipation rate $\kappa$.
\begin{figure}[ht]
\centering
\includegraphics[scale=0.60]{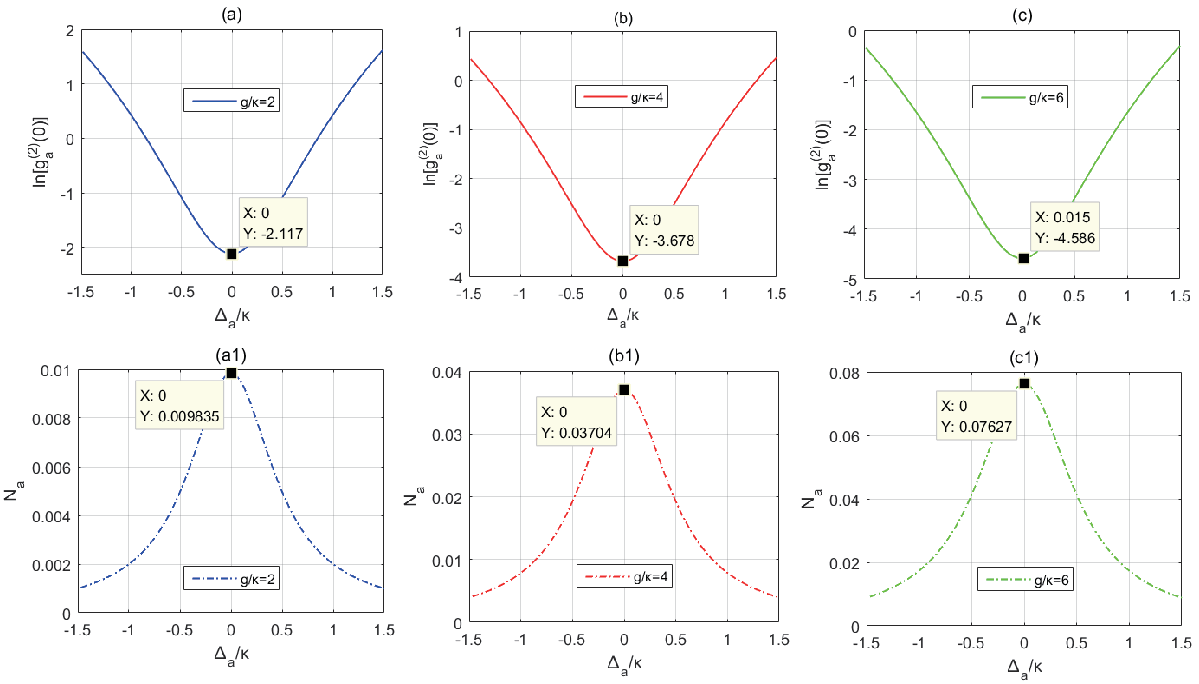}
\caption{(a) and (a1) logarithmic plot of $g^{(2)}(0)$ and average photon number $N_a$ as functions of $\Delta_a/\kappa$, with $F_a/\kappa=0.05$, $E/\kappa=0.01$, $\Delta_b/\kappa=-\Delta_c/\kappa=2$ and $g/\kappa=2$.  (b) and (b1) logarithmic plot of $g^{(2)}(0)$ and average photon number $N_a$ as functions of $\Delta_a/\kappa$, with $F_a/\kappa=0.05$, $E/\kappa=0.01$, $\Delta_b/\kappa=-\Delta_c/\kappa=2$ and $g/\kappa=4$. (c) and (c1) logarithmic plot of $g^{(2)}(0)$ and average photon number $N_a$ as functions of $\Delta_a/\kappa$, with $F_a/\kappa=0.05$, $E/\kappa=0.01$, $\Delta_b/\kappa=-\Delta_c/\kappa=2$ and \textbf{$g/\kappa=6$}.}
\label{fig3}
\end{figure}

In Figures. (a), (b), and (c), we logarithmically plot $g_{a}^{(2)}(0)$ as a function of $\Delta_a/\kappa$ for different values of the nonlinear coupling coefficient $g/\kappa$. The other parameters are set as $F_a/\kappa=0.05$, $E/\kappa=0.01$, and $\Delta_b/\kappa=-\Delta_c/\kappa=2$. From the computed results in the figure, it can be observed that the system can achieve single-photon blockade, and the minimum value of $g_{a}^{(2)}(0)$ occurs at $\Delta_a/\kappa=0$, consistent with the analytical conditions for achieving conventional photon blockade (CPB). Moreover, as the nonlinear coupling coefficient $g/\kappa$ increases, the bottom of the optimal blockade point becomes lower, indicating a more pronounced blockade effect. This observation aligns with the physical mechanism of CPB, which relies on a large nonlinear strength-induced level splitting in the system. Under the same parameter settings as in Figures. (a), (b), and (c), we plot the variation of the average photon number $N_b$ with $\Delta_a/\kappa$ in Figures. (a1), (b1), and (c1). By comparing the computational results, it is evident that the position of the optimal blockade point perfectly aligns with the location of the maximum average photon number. A higher average photon number indicates a higher brightness value of the system, and a higher brightness value in the optimal blockade region is more favorable for the system to achieve single-photon emission when implementing single-photon source devices.
\begin{figure}[ht]
\centering
\includegraphics[scale=0.60]{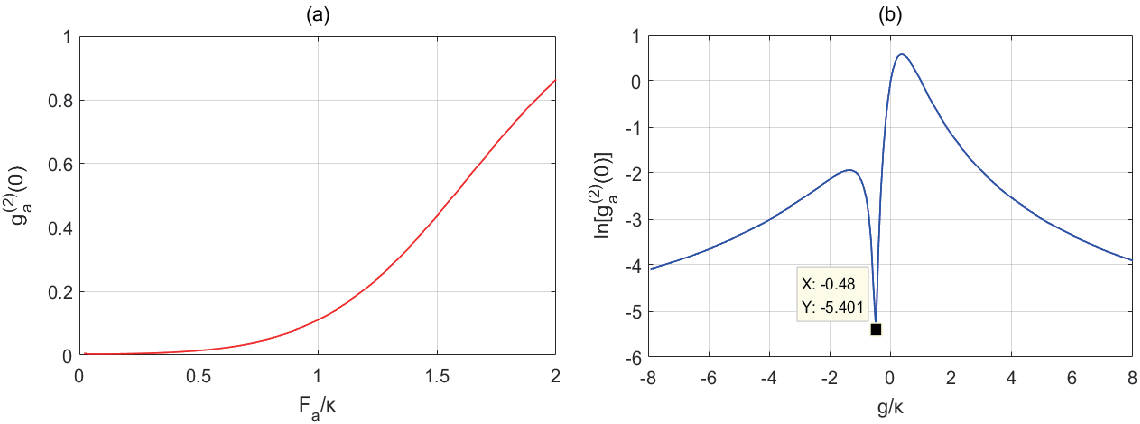}
\caption{(a) We plot $g_{a}^{(2)}(0)$ as a function of $F_a/\kappa$, where $g/\kappa=4$, $E/\kappa=0.01$, $\Delta_b/\kappa=-\Delta_c/\kappa=2$, and $\Delta_a/\kappa=0$. (b) Logarithmic plot of $g_{a}^{(2)}(0)$ as a function of $g/\kappa$, with $F_a/\kappa=0.05$, $E/\kappa=0.01$, $\Delta_b/\kappa=-\Delta_c/\kappa=2$, and $\Delta_a/\kappa=0$".}
\label{fig4}
\end{figure}

In the upcoming discussion, we'll delve into how the driving coefficient $F_a/\kappa$ and the nonlinearity coefficient $g/\kappa$ impact the implementation of conventional photon blockade (CPB) within the system. In Figure.(a), we depict $g^{(2)}(0)$ against $F_a/\kappa$, with $g/\kappa=4$, $E/\kappa=0.01$, $\Delta_b/\kappa=-\Delta_c/\kappa=2$, and $\Delta_a/\kappa=0$. The computational findings reveal that as $F_a/\kappa$ rises, the photon blockade effect gradually diminishes. Notably, when $F_a/\kappa>0.5$, the curve's slope significantly increases, and at $F_a/\kappa=2$, $g_{a}^{(2)}(0)$ approximates $1$, signaling the dissipation of the photon blockade effect. This arises from the Poissonian statistics of photons emitted by the driving source, contrasting the sub-Poissonian statistics essential for photon blockade. Hence, a high driving intensity for mode $a$ impedes achieving photon blockade. Generally, implementing conventional photon blockade necessitates meeting the weak driving condition. In Figure.\ref{fig4}(b), we display $g_{a}^{(2)}(0)$ against $g/\kappa$ logarithmically, with $F_a/\kappa=0.05$, $E/\kappa=0.01$, $\Delta_b/\kappa=-\Delta_c/\kappa=2$, and $\Delta_a/\kappa=0$. The computational analysis reveals that in the $g/\kappa>0$ region, increasing $g/\kappa$ leads to a gradual decrease in $g_{a}^{(2)}(0)$, indicating an augmented photon blockade effect, aligning with the physical mechanism of achieving conventional photon blockade. Conversely, in the $g/\kappa<0$ region, $g_{a}^{(2)}(0)$ decreases overall with increasing $g/\kappa$. However, at $g/\kappa=-0.48$, a noticeable dip emerges in the curve, signifying a substantial enhancement of the photon blockade effect. Drawing from prior analytical calculations, we ascertain that under the current parameter configuration, $g/\kappa=-0.48$ precisely fulfills the condition for implementing unconventional photon blockade (UPB) in the system (refer to Equation 9). Consequently, at $g/\kappa=-0.48$, the system exhibits UPB, corroborating the computational findings with theoretical predictions. Subsequently, we'll delve into the specifics of UPB within the system.

\subsection{Numerical Results of the system in implementing UPB}

\begin{figure}[ht]
\centering
\includegraphics[scale=0.60]{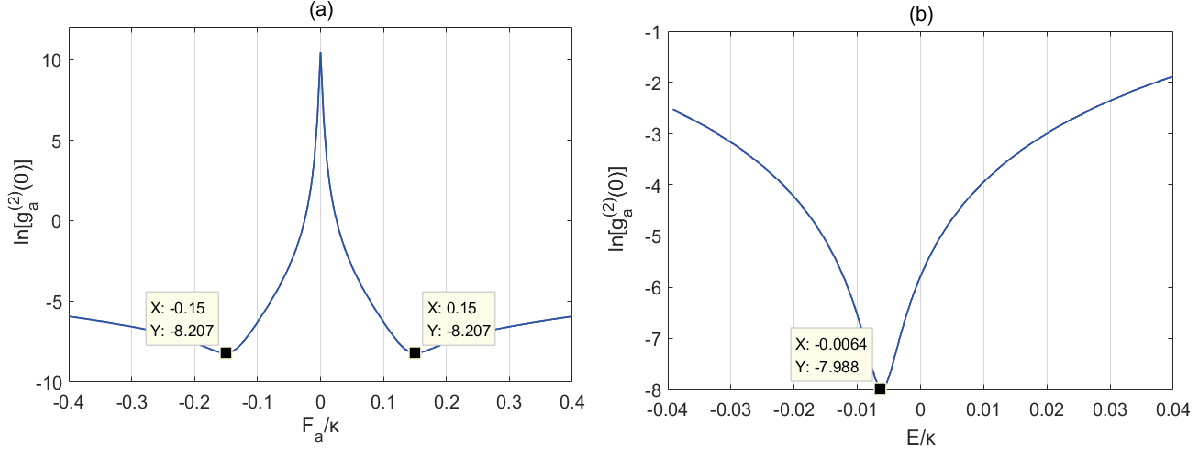}
\caption{(a) Logarithmic plot of $g^{(2)}(0)$ as a function of $F_a/\kappa$, with $g/\kappa=4$, $E/\kappa=0.01$, $\Delta_b/\kappa=0.5$, $\Delta_c/\kappa=1.5$, and $\Delta_a/\kappa=1$. (b) Logarithmic plot of $g^{(2)}(0)$ as a function of $E/\kappa$, with $g/\kappa=3$, $F_a/\kappa=0.1$, $\Delta_b/\kappa=0.5$, $\Delta_c/\kappa=1.5$, and $\Delta_a/\kappa=1.5$.}
\label{fig5}
\end{figure}

Next, we will discuss the numerical analysis of parameter settings required for implementing UPB in the system. In Figure~\ref{fig5}(a), we plot $g_{a}^{(2)}(0)$ as a function of $F_a/\kappa$ using logarithmic coordinates, while keeping $g/\kappa=4$, $E/\kappa=0.01$, $\Delta_b/\kappa=0.5$, $\Delta_c/\kappa=1.5$, and $\Delta_a/\kappa=1$. Here, we only satisfy the condition $2\Delta_a/\kappa=\Delta_b/\kappa+\Delta_c/\kappa$ for detuning, without considering the CPB analytical requirement of $\Delta_a/\kappa=0$. According to the computational results, the system exhibits single-photon blockade over a relatively wide range of $F_a/\kappa$, indicating that the driving coefficient of mode $a$ photons has a broad range of values for UPB implementation. Consistent with the optimal analytical solution for UPB implementation in the system, as shown in Equation 9, when $g/\kappa=4$ and the non-degenerate parameter amplification coefficient $E/\kappa=0.01$, the optimal blockade position in the system should occur at $F_a/\kappa=\pm0.15$, in line with theoretical predictions and numerical calculations. In Figure~\ref{fig5}(b), we plot $(g^{(2)}(0))$ as a function of $E/\kappa$, while keeping $F_a/\kappa=0.1$, $g/\kappa=3$, $\Delta_b/\kappa=1$, $\Delta_c/\kappa=2$, and $\Delta_a/\kappa=1.5$. Here, we also do not consider the CPB analytical requirement of $\Delta_a/\kappa=0$. The computational results demonstrate that the system can exhibit UPB. According to the optimal analytical solution for UPB implementation, when $g/\kappa=3$ and $F_a/\kappa=0.1$, the optimal blockade position in the system should occur at $E/\kappa=-0.0066$. The numerical solution aligns well with the analytical solution, albeit with a relatively narrow range of $E/\kappa$ values.

\begin{figure}[h]
\centering
\includegraphics[scale=0.62]{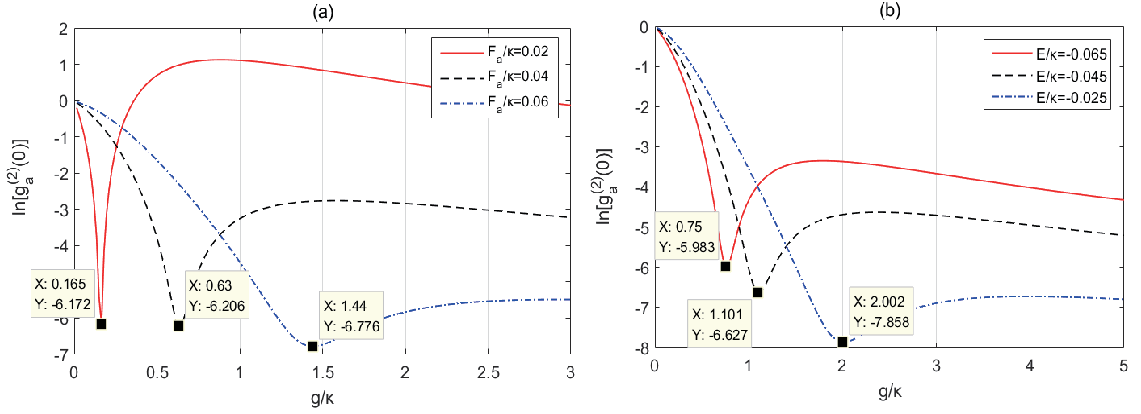}
\caption{(a) Logarithmic plot of $g_{a}^{(2)}(0)$ as a function of the nonlinear interaction strength $g/\kappa$, under different driving coefficients $F_{a}/\kappa$. Parameters used are $E/\kappa=-0.005$, $\Delta_{a}/\kappa=2$, and $\Delta_{a}/\kappa=\Delta_{b}/\kappa=1$. (b) Logarithmic plot of $g_{a}^{(2)}(0)$ as a function of the nonlinear interaction strength $g/\kappa$, under different nondegenerate parametric amplification factors $E/\kappa$. Parameters used are $F_{a}/\kappa=0.05$, $\Delta_{a}/\kappa=2$, and $\Delta_{a}/\kappa=\Delta_{b}/\kappa=1$.}
\label{fig6}
\end{figure}
In Figure.~\ref{fig6}(a), we plotted $g_{a}^{(2)}(0)$ as a function of $g/\kappa$ on a logarithmic scale for different values of $F_a/\kappa$. Here, we have $E/\kappa=-0.005$, $\Delta_b/\kappa=\Delta_c/\kappa=1$, and $\Delta_a/\kappa=2$. The computed results show that for the three different values of $F_a/\kappa$, the curves all exhibit a distinct dip, and the bottom values of the dip perfectly match the optimal blockade condition of the UPB. When $F_a/\kappa=0.02$, the optimal blockade point occurs at $g/\kappa=0.165$, indicating a relatively weak nonlinear strength in the system. This suggests that the current single-photon blockade arises from destructive quantum interference between different photon number states, consistent with the physical mechanism of UPB. In Figure.~\ref{fig6}(b), we plotted $g_{a}^{(2)}(0)$ as a function of $g/\kappa$ on a logarithmic scale for different values of $E/\kappa$. Here, we set $F_a/\kappa=0.05$, $\Delta_b/\kappa=\Delta_c/\kappa=1$, and $\Delta_a/\kappa=2$. Similar to Figure~.\ref{fig6}(a), we observe that single-photon blockade can occur in the region of small $g/\kappa$, and all three curves exhibit a distinct dip whose bottom values match the optimal blockade conditions of UPB. However, unlike Figure.~\ref{fig6}(a), in Figure.~\ref{fig6}(b), the blockade effect is significantly enhanced as $E/\kappa$ decreases. This indicates that smaller values of $E/\kappa$ are more favorable for the realization of UPB.

\subsection{Numerical Results of the system in implementing composite photon blockade}

\begin{figure}[ht]
\centering
\includegraphics[scale=0.60]{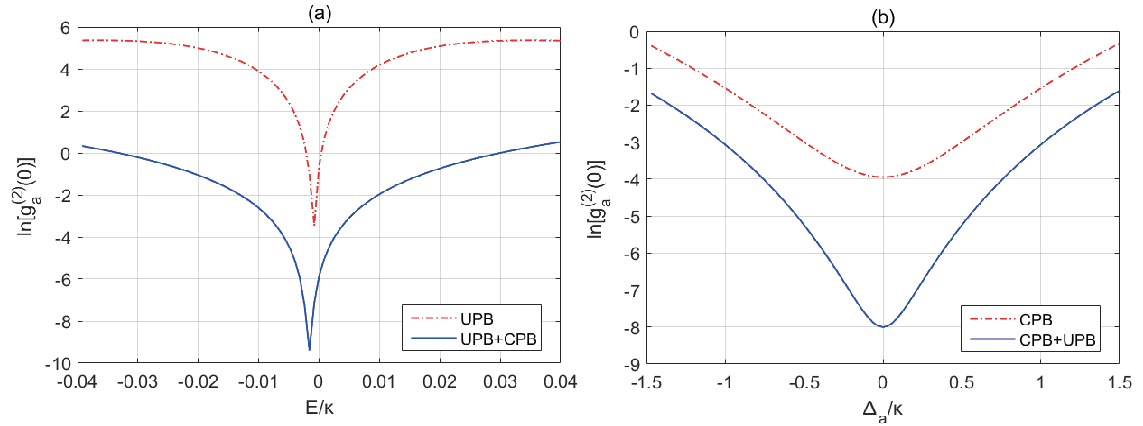}
\caption{(a) The logarithmic plot of $g_{a}^{(2)}(0)$ as a function of $E/\kappa$ is shown. The red dotted line corresponds to $F_a/\kappa=0.05$, $g/\kappa=3$, $\Delta_a/\kappa=2$, $\Delta_b/\kappa=1$, and $\Delta_c/\kappa=1$. The blue solid line corresponds to $F_a/\kappa=0.05$, $g/\kappa=3$, $\Delta_b/\kappa=-\Delta_c/\kappa=2$, and $\Delta_a/\kappa=0$. (b) The logarithmic plot of $g^{(2)}(0)$ as a function of $\Delta_a/\kappa$ is shown. The red dotted line corresponds to $F_a/\kappa=0.1$, $g/\kappa=3$, $\Delta_b/\kappa=-\Delta_c/\kappa=3$, and $E/\kappa=0.01$. The blue solid line corresponds to $F_a/\kappa=0.1$, $g/\kappa=3$, $\Delta_b/\kappa=-\Delta_c/\kappa=3$, and $E/\kappa=-0.0066$.}
\label{fig7}
\end{figure}
\begin{figure}[ht]
\centering
\includegraphics[scale=0.60]{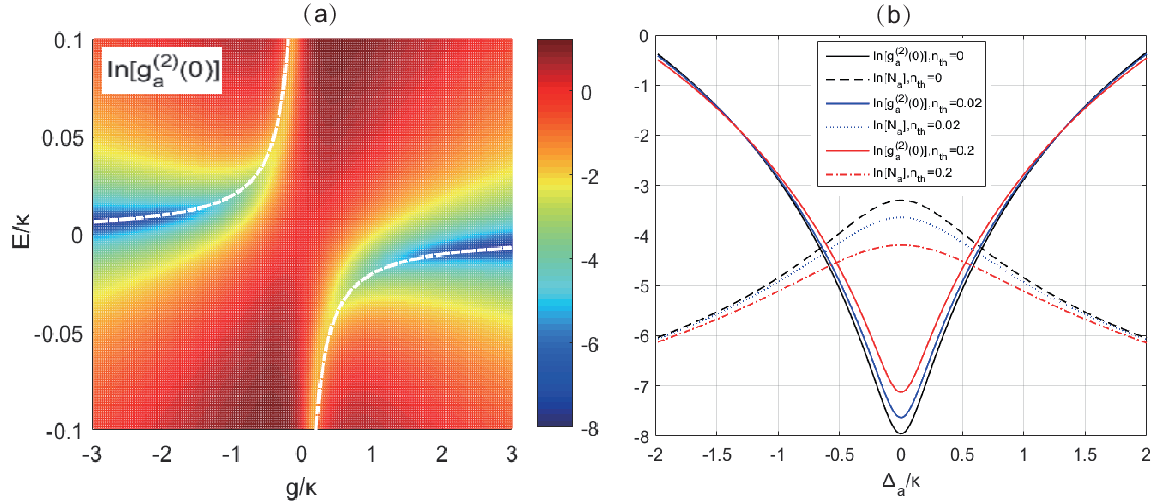}
\caption{(a) Logarithmic plot of $g^{(2)}(0)$ as a function of $g/\kappa$ and $E/\kappa$ for $F_a/\kappa=0.1$, $\Delta_b/\kappa=-\Delta_c/\kappa=2$, and $\Delta_a/\kappa=0$. (b) Logarithmic plot of $g_{a}^{(2)}(0)$ and $N_{a}$ as a function of $\Delta_a/\kappa$ under different numbers of thermal photons $n_{th}$. Parameters used are $F_{a}/\kappa=0.1$, $E/\kappa=-0.0066$, $\Delta_{b}/\kappa=-\Delta_{c}/\kappa=2$, and $g/\kappa=3$.}
\label{fig8}
\end{figure}

In the previous discussion, we separately investigated the implementation of CPB and UPB in the current system. According to previous research results, we know that when CPB is achieved, the value of $N_a$ is large, resulting in high brightness, but the value of the second-order correlation function is also large, indicating low photon purity. On the other hand, when UPB is achieved, the second-order correlation function has a small value, indicating high photon purity, but the value of $N_a$ is small, resulting in low brightness. Therefore, the implementation of a composite photon blockade system that allows CPB and UPB to occur simultaneously within the same parameter range is meaningful for achieving higher purity and brightness in a single-photon source. In Figure.7(a), the red dashed line represents the plot of $g_a^{(2)}(0)$ as a function of $E/\kappa$, using a logarithmic scale. The parameters are set as $\Delta_b/\kappa=\Delta_c/\kappa=1$, $\Delta_a/\kappa=2$, and $F_a/\kappa=0.05$. The computed results show that the system achieves single-photon blockade, and the optimal blockade point perfectly matches the analytical solution of UPB as given by Equation 9. At this point, $\Delta_a/\kappa=2$, and the CPB condition is not satisfied. This indicates that in the calculation results represented by the red dashed line, only UPB occurs. In the blue curve, we similarly plot $g_a^{(2)}(0)$ as a function of $E/\kappa$, using a logarithmic scale. The parameters are set as $\Delta_b/\kappa=-\Delta_c/\kappa=1$, $\Delta_a/\kappa=0$, and $F_a/\kappa=0.05$. The computed results show that the optimal blockade point coincides with the red dashed line, and the depth of the curve's minimum is lower, indicating a stronger blockade effect. In the calculation results represented by the blue curve, the condition of $\Delta_a/\kappa=0$ satisfies CPB. The optimal blockade point also perfectly matches the analytical solution of UPB as given by Equation 9. Therefore, CPB and UPB coexist in this case, which is referred to as composite photon blockade. To further discuss the implementation of composite photon blockade in the system, in Figure. 7(b), we plot $g_a^{(2)}(0)$ as a function of $\Delta_a/\kappa$ using a logarithmic scale. In the red dashed line, we set $g/\kappa=3$, $\Delta_b/\kappa=-\Delta_c/\kappa=3$, $E/\kappa=0.01$, and $F_a/\kappa=0.1$. The computed results show that the system achieves single-photon blockade, and the optimal blockade point occurs at $\Delta_a/\kappa=0$, consistent with the analytical solution of CPB. In the blue curve, we set $F_a/\kappa=0.1$, $g/\kappa=3$, and $E/\kappa=-0.0066$ to satisfy the analytical conditions of UPB as shown in Equation 9. The system exhibits a more pronounced photon blockade effect at $\Delta_a/\kappa=0$. In other words, in the calculation results represented by the blue curve, the parameter settings satisfy the analytical solution of UPB, and the optimal blockade point perfectly matches the analytical solution of CPB. Therefore, CPB and UPB coexist in this case, which is referred to as composite photon blockade. Therefore, the computational results in Figures. 7 demonstrate that the current system can achieve composite photon blockade, and the implementation of composite photon blockade has significant purity advantages compared to UPB and CPB individually.

To further demonstrate the simultaneous implementation of CPB and UPB in the system within the same parameter range, we present Figure.8(a), where we plot $g_{a}^{(2)}(0)$ as a function of $g/\kappa$ and $E/\kappa$ using a logarithmic scale. Here, $F_a/\kappa=0.1$, $\Delta_b/\kappa=-\Delta_c/\kappa=2$, and $\Delta_a/\kappa=0$. The white dashed line represents the optimal analytical solution for achieving UPB, as given by Equation (9). The computed results show that the region of photon blockade aligns with the analytical solution for UPB. To investigate the influence of environmental temperature on the system, we present Figure. 8(b), where we plot $g_{a}^{(2)}(0)$ and $N_{a}$ as functions of different thermal photon numbers (${n}_{th}$). The parameters are set as $F_a/\kappa=0.1$, $E/\kappa=-0.0066$, $\Delta_b/\kappa=-\Delta_c/\kappa=2$, and $g/\kappa=3$. The computed results indicate that as ${n}_{th}$ increases, the value of $g_{a}^{(2)}(0)$ tends to increase, while the average photon number $N_{a}$ decreases. This suggests that lower temperatures are favorable for achieving photon blockade.

\subsection{Numerical Results of the system in implementing two photon blockade}

\begin{figure}[ht]
\centering
\includegraphics[scale=0.60]{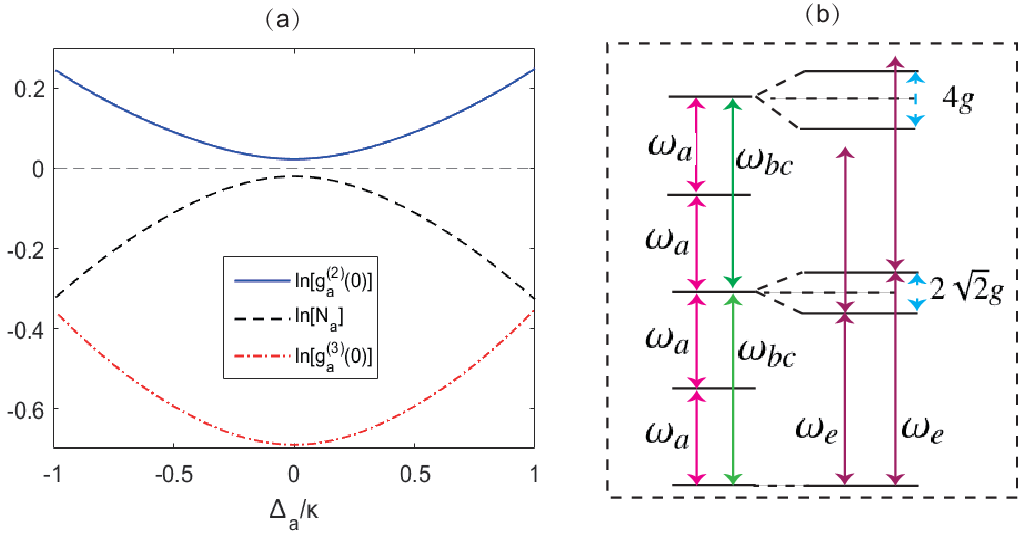}
\caption{(a) Logarithmic plot of $g_{a}^{(2)}(0)$, $g_{a}^{(3)}(0)$, and $N_{a}$ as a function of $\Delta_a/\kappa$ for $E/\kappa=0.06$, $\Delta_b/\kappa=-\Delta_c/\kappa=2$, $g/\kappa=5$, and $F_a/\kappa=0$. (b) The energy level schematic diagram illustrating the realization of two-photon bunching (2PB) in the system.}
\label{fig9}
\end{figure}
Finally, we will discuss the case of achieving two-photon blockade (2PB) in the system. 2PB, as one of the primary methods for realizing sub-Poissonian light sources, has garnered significant attention. Implementing 2PB requires satisfying the conditions of both three-photon antibunching and two-photon bunching simultaneously, i.e., $g^{(2)}(0) \geq 1$ and $g^{(3)}(0) < 1$ must hold concurrently. In Figure. 9(a), we present the logarithmic plot of $g_{a}^{(2)}(0)$, $g_{a}^{(3)}(0)$, and $N_{a}$ as a function of $\Delta_a/\kappa$, where $F_a/\kappa=0$, $E/\kappa=0.06$, $\Delta_{b}/\kappa=-\Delta_{c}/\kappa=2$, and $g/\kappa=5$. Computational results demonstrate that the system can achieve 2PB in the region of $\Delta_a/\kappa=0$, consistent with the predicted outcomes. In Figure. 9(b), we provide a schematic energy level diagram illustrating how the system accomplishes 2PB in mode $a$. The diagram depicts multiple energy levels representing the quantum states of mode $a$. The bottom-most level represents the ground state, while the levels above it correspond to excited states. An energy difference typically denoted as $\omega_a$ exists between these two levels, and $\omega_{bc}$ represents the amplification frequency of the non-degenerate parametric drive. According to the energy level diagram, it is evident that when the driving frequencies of modes b and c photons are set to $\omega_{bc}=\omega_{e}$, the system can achieve 2PB in mode $a$, with 2PB occurring in the region of $\Delta_a/\kappa=\pm1$.

\section{Conclusion}

We investigated the phenomenon of photon blockade (PB) in a non-degenerate four-wave mixing system. Through analytical calculations, we obtained analytical solutions for implementing UPB and CPB, which were validated by numerical solutions obtained by solving the master equation in the steady-state limit. We conducted comprehensive numerical analysis and discussions on the system parameters required for implementing CPB and UPB. It is noteworthy that when CPB and UPB occur simultaneously within the same parameter range, their  $g_{a}^{(2)}(0)$ values are lower than those when CPB or UPB occurs alone. We refer to this phenomenon of CPB and UPB occurring simultaneously within the same parameter range as the composite photon blockade effect. Implementing composite photon blockade provides an effective solution to address issues such as low single-photon purity and high nonlinearity requirements in CPB implementation, as well as poor single-photon brightness in UPB implementation. Finally, we conducted theoretical analysis and numerical calculations for the case of two-photon blockade(2PB) in the system, and the results demonstrate that 2PB can occur when the system meets the conditions for single-driving mode.

\section*{Acknowledgment}
This work is supported by the Jilin Provincial Natural Science Foundation, Grant No. 20240101305JC. JiLin Education Department Fund under Grant No. JJKH20230015KJ. Program for Innovative Research Team of Baicheng Normal University, Grant No.IRTBCNU.

\section*{Disclosures:} The authors declare no conflicts of interest.


\bibliography{mybibfile.bib}

\end{document}